\RequirePackage{fix-cm}

\documentclass[smallcondensed]{svjour3}       

\smartqed  
\usepackage{graphicx}
\usepackage{natbib}
\usepackage[]{units}

\begin{document}

\title{Reflecting about Selecting Noninformative Priors
}


\author{Kaniav Kamary         \and
        Christian. P. Robert 
}


\institute{K. Kamary \at
              CEREMADE, Universit\'e Paris-Dauphine, 75775 Paris cedex 16, France \\
              \email{kamary@ceremade.dauphine.fr}           
           \and
           C. P. Robert \at
              CREST-Insee and CEREMADE, Universit\'e Paris-Dauphine, 75775 Paris cedex 16, France\\
              Dept. of Statistics, University of Warwick \email{xian@ceremade.dauphine.fr} 
}


\maketitle

\begin{abstract}
Following the critical review of \citet{Hd}, we reflect on what is presumably
the most essential aspect of Bayesian statistics, namely the selection of a
prior density.  In some cases, Bayesian inference remains fairly stable under a
large range of noninformative prior distributions. However, as discussed by \citet{Hd}, 
there may also be unintended consequences of a choice of a noninformative prior and,
these authors consider this problem ignored in Bayesian studies. As they based
their argumentation on four examples, we reassess these examples and their
Bayesian processing via different prior choices. Our conclusion is to 
lower the degree of worry about the impact of the prior, exhibiting an 
overall stability of the posterior distributions. We thus consider that
the warnings of \citet{Hd}, while commendable, do not jeopardize the use of
most noninformative priors. 

\end{abstract}

\section{Introduction}\label{intro}

The choice of a particular prior for the Bayesian analysis of a statistical
model is often seen more as an art than as a science. When the prior cannot be
derived from the available information, it is generaly constructed as a
noninformative prior. This derivation is mostly mathematical and, even though
the corresponding posterior distribution has to be proper and hence constitutes
a correct probability density, it nonetheless leaves the door open to
criticism. The focus of this note is the paper by \citet{Hd}, where the authors
consider using a particular noninformative distribution as a problem in itself,
often bypassed by users of these priors: ``if parameters with diffuse proper
priors are subsequently transformed, the resulting induced priors can, of
course, be far from diffuse, possibly resulting in unintended influence on the
posterior of the transformed parameters'' (p.77). Using the inexact argument that
most problems rely on MCMC methods and {\em hence} require proper priors, the
authors restrict the focus to those priors.

In their critical study, \citet{Hd} investigate the negative side effects of
some specific prior choices related with specific examples. Our note aims at
re-examining their investigation and at providing a more balanced discussion on
these side effects. We first stress that a prior is considered as {\em
informative} by \citet{Hd} ``to the degree it renders some values of the
quantity of interest more likely than others'' (p.77), and with this
definition, when comparing two priors, the prior that is more informative is
deemed preferable. In contrast with this definition, we consider that an {\em
informative} prior expresses specific, definite (prior) information about the
parameter, providing quantitative information that is crucial to the estimation
of a model through restrictions on the prior distribution \citep{Bc}.  However,
in most practical cases, a model parameter has no substance {\em per se} but
instead calibrates the probability law of the random phenomenon observed
therein. The prior is thus a tool that summarizes the information available on
this phenomenon, as well as the uncertainty within the Bayesian structure. Many
discussions can be found in the literature on how appropriate choices between
the prior distributions can be decided. In this case, robustness considerations
also have an important role to play \citep{Cp,Psa}. This point of view will be
obvious in this note as, e.g., in processing a logistic model in the following
section.  Within the sole setting of the examples first processed in
\citet{Hd}, we do exhibit a greater stability in the posterior distributions
through various noninformative priors. 

The plan of the note is as follows: we first provide a brief review of
noninformative priors in Section \ref{sec:zero}.  In Section \ref{sec:un}, we
propose a Bayesian analysis of a logistic model (Seaman III et al.'s (2012)
first example) by choosing the normal distribution $N(0, \sigma^2)$ as the
regression coefficient prior. We then compare it with a $g$-prior, as well as
flat and Jeffreys' priors, concluding to the stability of our results. The next
sections cover the second to fourth examples of \citet{Hd}, modeling covariance
matrices, treatment effect in biomedical studies, and a multinomial
distribution. When modeling covariance matrices, we compare two default priors
for the standard deviations of the model coefficients.  In the multinomial
setting, we discuss the hyperparameters of a Dirichlet prior.  Finally, we
conclude with the argument that the use of noninformative priors is reasonable
within a fair range and that they provide efficient Bayesian estimations when
the information about the parameter is vague or very poor. 

\section{Noninformative priors}\label{sec:zero}

As mentioned above, when prior information is unavailable and if we stick to
Bayesian analysis, we need to resort to one of the so-called {\em
noninformative priors}. Since we aim at a prior with minimal impact on the
final inference, we define a noninformative prior as a statistical distribution
that expresses vague or general information about the parameter in which we are
interested. In constructive terms, the first rule for determining a
noninformative prior is the principle of indifference, using uniform
distributions which assign equal probabilities to all possibilities
\citep{Lpl}. This distribution is however not invariant under reparametrization
\citep[see][for references]{Br,Bc}. If the problem does not allow for an
invariance structure, Jeffreys' (\citeyear{Jeff}) priors, then reference
priors, exploit the probabilistic structure of the problem under study in a
more formalised way.  Other methods have been advanced, like the little-known
data-translated likelihood of \citet{Bto}, maxent priors \citep{Jaynes},
minimum description length priors \citep{Riss} and probability matching priors
\citep{Wh}.

\citet{Bt} envision noninformative priors as a mere mathematical tool, while
accepting their feature of minimizing the impact of the prior selection on
inference: ``Put bluntly, data cannot ever speak entirely for themselves, every
prior specification has some informative posterior or predictive implications
and \textit{vague} is itself much too vague an idea to be useful. There is no
``objective" prior that represents ignorance'' (p.298). There is little to
object against this quote since, indeed, prior distributions can never be
quantified or elicited exactly, especially when no information is available on
those parameters. Hence, the concept of ``true" prior is meaningless and the
quantification of prior beliefs operates under uncertainty. As stressed by
\citet{Rbv},  noninformative priors enjoy the advantage that they can be
considered to provide robust solutions to relevant problems even though ``the
user of these priors should be concerned with robustness with respect to the
class of reasonable noninformative priors'' (p.59).

\section{Example 1: Bayesian analysis of the logistic model}\label{sec:un}

The first example in \citet{Hd} is a standard logistic regression modelling
the probability of coronary heart disease as dependent on the age $x$ by
\begin{equation} 
\rho(x)=\frac{\mathrm{exp}{(\alpha +\beta x)}}{1+\mathrm{exp}{(\alpha +\beta x)}}.
\label{2}
\end{equation}
First we recall the original discussion in \citet{Hd} and then run our
own analysis by selecting some normal priors as well as the
$g$-prior, the flat prior and Jeffreys' prior.

\subsection{Seaman et al.'s (2012) analysis}

\begin{figure}[!h]
\centerline{
\includegraphics[scale=0.6,width=0.65\textwidth]{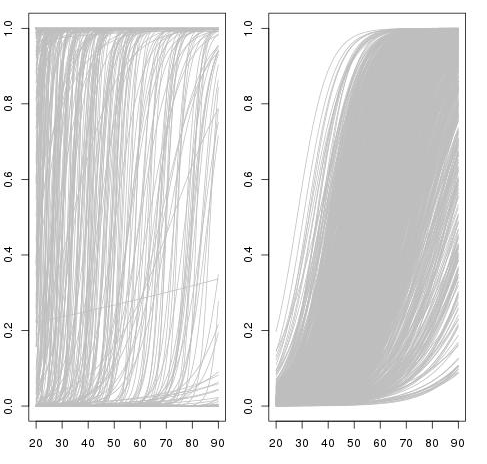}}
\caption{\small Logistic cdfs across a few thousans simulations from the 
normal prior, when using the prior selected by \citet{Hd} (left) and the 
prior defined as the $G$-prior(right)}
\label{RB}
\end{figure}

For both parameters of the model (\ref{2}), \citet{Hd} chose a normal prior
$N(0,\sigma^2)$. A first surprising feature in this choice is to opt for an
{\em identical} prior on both intercept and slope coefficients, instead of,
e.g., a $g$-prior (discussed in the following) that would rescale each
coefficient according to the variation of the corresponding covariate. Indeed,
since $x$ corresponds to age, the second term $\beta x$ in the regression
varies $50$ times more than the intercept. When plotting logistic cdf's induced
by a few thousands simulations from the prior, those cumulative functions
mostly end up as constant functions with the extreme values $0$ and $1$. This
behaviour is obviously not particularly realistic since the predicted
phenomenon is the occurrence of coronary heart disease. Under this minimal
amount of information, the prior is thus using the wrong scale: the simulated
cdfs should have a reasonable behavior over the range $(20,100)$ of the
covariate $x$. For instance, it should focus on a $-5$ log-odds ratio at age
$20$ and a $+5$ log-odds ratio at $100$, leading to the comparison pictured in
Figure \ref{RB} (left versus right). Furthermore, the fact that the coefficient
of $x$ may be negative also bypasses a basic item of information about the
model and answers the later self-criticism in \citet{Hd} that the prior
probability that the ED50 is negative is $0.5$. Using instead a flat prior
would answer the authors' criticisms about the prior behavior, as we now
demonstrate.

We stress that \citet{Hd} produce no further justification for the choice of the prior
variance $\sigma^2=25^2$, other than there is no information about the model
parameters. This is a completely arbitrary choice of prior, arbitrariness that
does have a considerable impact on the resulting inference, as already discussed.
\citet{Hd} further criticized the chosen prior by comparing
both posterior mode and posterior mean derived from the normal prior assumption
with the MLE. If the MLE is the golden standard there then one may wonder about
the relevance of a Bayesian analysis! When the sample size $N$
gets large, most simple Bayesian analyses based on noninformative prior
distributions give results similar to standard non-Bayesian approaches
\citep{Bad}. For instance, we can often interpret classical point estimates as
exact or approximate posterior summaries based on some implicit full
probability model. Therefore, as $N$ increases, the influence of the prior on
posterior inferences decreases and, when $N$ goes to infinity, most priors lead
to the same inference.  However, for smaller sample sizes, it is
inappropriate to summarize inference about the parameter by one value like the
mode or the mean, especially when the posterior distribution of the parameter
is more variable or even asymmetric.

\begin{figure}[!h]
\centerline{
\includegraphics[width=0.5\textwidth]{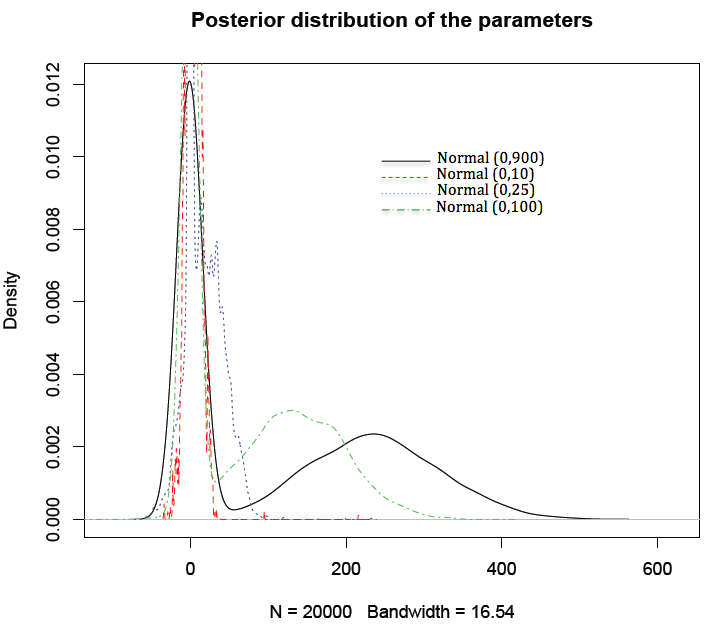}}
\caption{\small 
Posterior distributions of the logistic parameter $\alpha$  when priors are $N(0,
\sigma)$ for $\sigma = 10, 25, 100, 900$, based on $10^4$ MCMC simulations.
\label{NP}}
\end{figure}

The dataset used here to infer on $(\alpha, \beta)$ is the Swiss
banknote benchmark (available in R). The response variable $y$
indicates  the state of the banknote, i.e. whether the bank note is genuine or
counterfeit. The explanatory variable is the bill length.  This data yields
the maximum likelihood estimates $\tilde{\alpha}=233.26$ and
$\tilde{\beta}=-1.09$. To check the impact of the normal prior variance, we
used a random walk Metropolis-Hastings algorithm as in \cite{Bco} and derived
the estimators reproduced in Table \ref{j10}.  We can spot definitive changes
in the results that are caused by moves in the coefficient $\sigma$, hence
concluding to the clear sensitivity of the posterior to the choice of
hyperparameter $\sigma$ (see also Figure  \ref{NP}).

\begin{table}
  \caption{Posterior estimates of the logistic parameters 
using a normal prior when $\sigma=10, 25, 100, 900$}
  \label{j10} 
\centerline{
\begin{tabular}{|c|c|c|c|}
\hline
\multicolumn{4}{|c|}{$\sigma=10$} \\
\hline
\multicolumn{2}{|c|}{ $\hat{\alpha}$} & \multicolumn{2}{|c|}{$\hat{\beta}$} \\
\hline  
 mean  & s.d           &  mean   & s.d \\ \cline{1-4}
  3.482 & 11.6554 & -0.0161 & 0.0541 \\
\hline
\multicolumn{4}{|c|}{$\sigma=25$} \\
\hline
18.969 & 24.119 & -0.0882 & 0.1127 \\
\hline
\multicolumn{4}{|c|}{$\sigma=100$} \\
\hline
137.63 & 64.87 &  -0.6404 & 0.3019\\
\hline
\multicolumn{4}{|c|}{$\sigma=900$} \\
\hline
 237.2 & 86.12 & -1.106 & 0.401\\
\hline
\end{tabular}}
\end{table}

\subsection{Larger classes of priors}

Normal priors are well-know for their lack of robustness (see e.g.
\citealp{Rbv}) and the previous section demonstrates the long-term
impact of $\sigma$. However, we can limit variations in the posteriors, using the
$g$-priors of \citet{Zr},
\begin{equation} 
\alpha, \beta \mid X \sim N_{2}(0, g(X^T X)^{-1}).  \label{3}
\end{equation}
where the prior variance-covariane matrix is a scalar multiple 
of the information matrix for the linear regression. This coefficient
$g$ plays a decisive role in the analysis, however large values of $g$ imply a more diffuse prior
and, as shown e.g. in \citet{Bco}, if the value of $g$ is large enough, the Bayes estimate stabilizes.
We will select $g$ as equal to the sample size $200$, following \citet{Mgp}, as it means that the
amount of information about the parameter is equal to the amount of information contained in one single observation.

A second reference prior is the flat prior $\pi(\alpha, \beta)=1$. And Jeffreys' prior constitutes our
third prior as in \cite{Bco}. In the logistic case, Fisher's information matrix is
$\mathbf{I}(\alpha, \beta, X)=X^TWX$, where $X=\{x_{ir}\}$ is the design matrix, 
$W=\mathrm{diag}\{m_i \pi_i (1-\pi_i)\}$ 
and $m_i$ is the binomial index for the ith count \citep{Bme}. This leads to Jeffreys' prior
$\{\mathrm{det}(\mathbf{I}(\alpha, \beta, X))\}^{\nicefrac{1}{2}}$, proportional to
$$
 \left[ \sum_{i=1}^n \frac{\mathrm{exp}{(\alpha+\beta x_i)}}{\{1+\mathrm{exp}{(\alpha+\beta x_i)}\}^2}\sum_{i=1}^n \frac{x_i^2 \mathrm{exp}{(\alpha+\beta x_i)}}{\{1+\mathrm{exp}{(\alpha+\beta x_i)}\}^2}-\left\{ \sum_{i=1}^n \frac{x_i \mathrm{exp}{(\alpha+\beta x_i)}}{\{1+\mathrm{exp}{(\alpha+\beta x_i)}\}^2} \right\}^2\right]^{\frac{1}{2}}
$$
This is a nonstandard distribution on $(\alpha, \beta)$ but it can be easily approximated by 
a Metropolis-Hastings algorithm whose proposal is the normal Fisher approximation of the likelihood, as in \cite{Bco}.

\begin{figure}[!h]
\centerline{
\includegraphics[scale=1,width=1\textwidth]{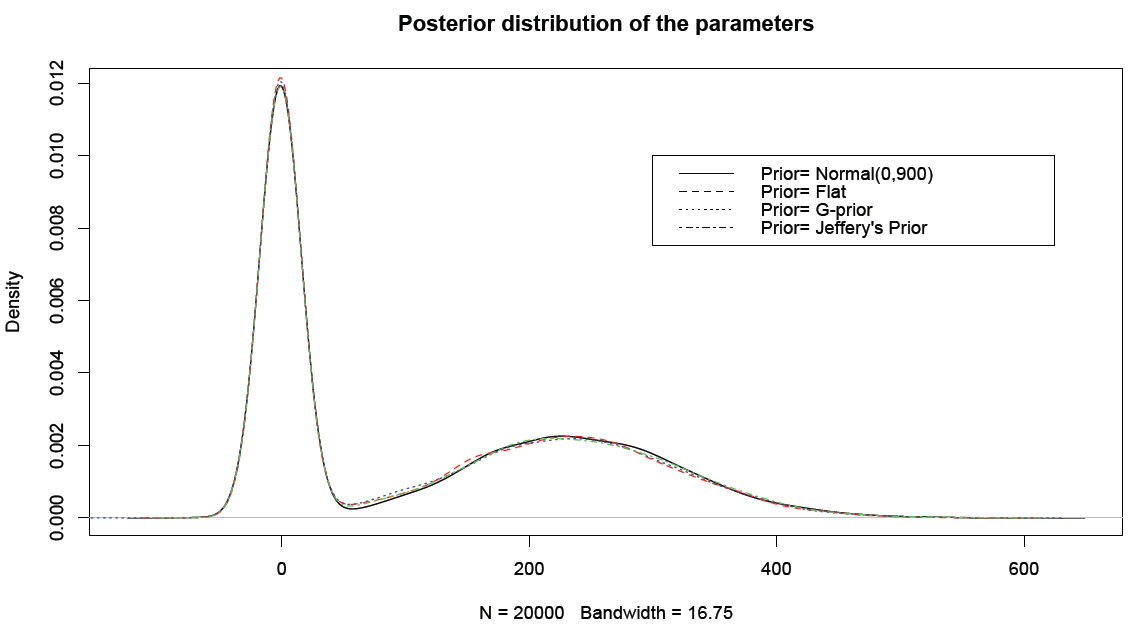}}
\caption{\small Posterior distributions of the parameters of the logistic model  when the prior is $N(0, 900^2)$, g-prior, flat prior and Jeffreys' prior, respectively. The estimated posterior distributions are based on $10^4$ MCMC iterations.}
\label{MP}
\end{figure}

Bayesian estimates of the regression coefficients associated with the above three noninformative priors 
are summarized in Table \ref{j2}. Those estimates vary quite moderately from one choice to the next, as well
as relatively to the MLEs and to the results shown in Table \ref{j10} when $\sigma=900$. Figure
\ref{MP} is even more definitive about this stability of Bayesian inferences under different 
noninformative prior choices.

\begin{table}
  \caption{Posterior estimates of the logistic parameters
under a $g$-prior, a flat prior and Jeffreys' prior for the banknote benchmark. 
Posterior means and standard deviations remain quite similar under all priors.
All point estimates are averages of MCMC samples of size $10^4$.} 
  \label{j2} 
\centerline{
\begin{tabular}{|c|c|c|c|}
\hline
\multicolumn{4}{|c|}{$g$-prior} \\
\hline  
\multicolumn{2}{|c|}{ $\hat{\alpha}$} & \multicolumn{2}{|c|}{ $\hat{\beta}$} \\
\hline 
mean & s.d & mean & s.d \\ \cline{1-4}
237.63 & 88.0377 & -1.1058 & 0.4097\\
\hline
\multicolumn{4}{|c|}{Flat prior} \\
\hline
 236.44 & 85.1049 & -1.1003 & 0.3960 \\
\hline
\multicolumn{4}{|c|}{Jeffreys' prior} \\
\hline
237.24 & 87.0597 & -1.1040 & 0.4051\\
\hline
\end{tabular}}
\end{table}

\vspace{0.3cm}
\section{Example 2: Modeling covariance matrices}
\label{sec:six}

The second choice of prior criticized by \citet{Hd}, was proposed by
\citet{Bmm} for the modeling of covariance matrices. However the paper falls
short of demonstrating a clear impact of this prior modelling on posterior
inference.  Furthermore the adeopted solution of using another proper prior resulting in
a ``wider" dispersion requires a prior knowledge of how wide is wide enough. We
thus run Bayesian analyses considering prior beliefs specified by both \citet{Hd} and \citet{Bmm}.

\subsection{Setting}
The multivariate regression model of \citet{Bmm} is 

\begin{equation}
Y_j \mid X_j, \beta_j, \tau_j \sim N(X_j \beta_j, \tau_j^2 I_{n_j}), ~~j=1, 2, \ldots, m.
\label{e1}
\end{equation}
where $Y_j $ is a vector of $n_j$ dependent variables, $X_j$ is an $n_j \times
k$ matrix of covariate variables, and $ \beta_j$ is a $k$-dimensional
parameter vector. For this model, \citet{Bmm} considered an iid 
normal distribution as the prior 
$$
\beta_j \sim N(\bar{\beta}, \Sigma)
$$ 
conditional on $\bar{\beta}, \Sigma$ where $ \bar{\beta}, \tau_j^2$
for $j=1, 2,\ldots, m$ are independent and follow a normal and inverse-gamma
priors, respectively. Assuming that $\bar{\beta}, \tau_j^2$'s and $\Sigma$ are
a priori independent, \citet{Bmm} firstly provide a full discussion on how to
choose a prior for $\Sigma$ because it determines the nature of the shrinkage
of the posterior of the individual $ \beta_j$ is towards a common target. 
The covariance matrix $\Sigma$ is defined as a diagonal matrix with diagonal elements $S$, multiplied by
a $k \times k$ correlation matrix $R$, 
$$
\Sigma=\mathrm{diag}(S) R\mathrm{diag}(S)\,.
$$
Note that $S$ is the $k \times 1$ vector of standard deviations of the $ \beta_j$s, $(S_1, \ldots, S_k)$. \citet{Bmm}
propose lognormal distributions as priors on $
S_j$. The correlation matrix could have (1) a joint uniform prior $p(R) \propto 1$, or (2) a marginal
prior obtained from the inverse-Wishart distribution for $\Sigma$ which means $p(R)$ is derived from 
the integral over $S_1, \ldots, S_k$ of a standard inverse-Wishart distribution. In the second case, 
all the marginal densities for $r_{ij}$ are uniform when $i\neq j$ \citep{Bmm}.

Considering the case of a single regressor, i.e. $k=2$,
\citet{Hd} chose a different prior structure, with a flat prior on the
correlations and a lognormal prior with means $1$ and $-1$, and standard deviations
$1$ and $0.5$ on the standard deviations of the intercept and slope, respectively.
Simulating from this prior, they concluded at a high concentration near zero.
They then suggested that the lognormal distribution should be replaced by a
gamma distribution $G(4, 1)$ as it implies a more diffuse prior. 
The main question here is whether
or not the induced prior is more diffuse should make us prefer gamma to
lognormal as a prior for $S_j$, as discussed below.

\subsection{Prior beliefs} 

First, Barnard et al.'s (2000) basic modeling intuition is ``that each
regression is a particular instance of the same type of relationship" (p.1292).
This means an exchangeabile prior belief on the regression parameters.  As an
example, they suppose that  $m$ regressions are similar models where each
regression corresponds to a different firm in the same industry branch. Exploiting
this assumption, when $ \beta_j$ has a normal prior like $\beta_{ij} \sim N(
\bar{\beta}_i,  \sigma_i^2),\ j=1, 2, \ldots, m$, the standard deviation of $
\beta_{ij} $ ($S_i= \sigma_i$) should be small as well so ``that the
coefficient for the $i$th explanatory variable is similar in the different
regressions" (p.1293). In other words, $S_i$ concentrated on small values
implies little variation in the $i$th coefficient. Toward this goal,
\citet{Bmm} chose a prior concentrated close to zero for the standard deviation
of the slope so that the posterior of this coefficient would be shrunken
together across the regressions.  Based on this basic idea and taking tight
priors on $\Sigma$ for $\beta_j, j=1,\ldots, m$, they investigated the
shrinkage of the posterior on $\beta_j$ as well as the degree of similarity of
the slopes.  Their analysis showed that a standard deviation prior that is more
concentrated on small values results in substantial shrinkage in the
coefficients relative to other prior choices.

\begin{figure}[!h]
\centerline{
\includegraphics[scale=0.5]{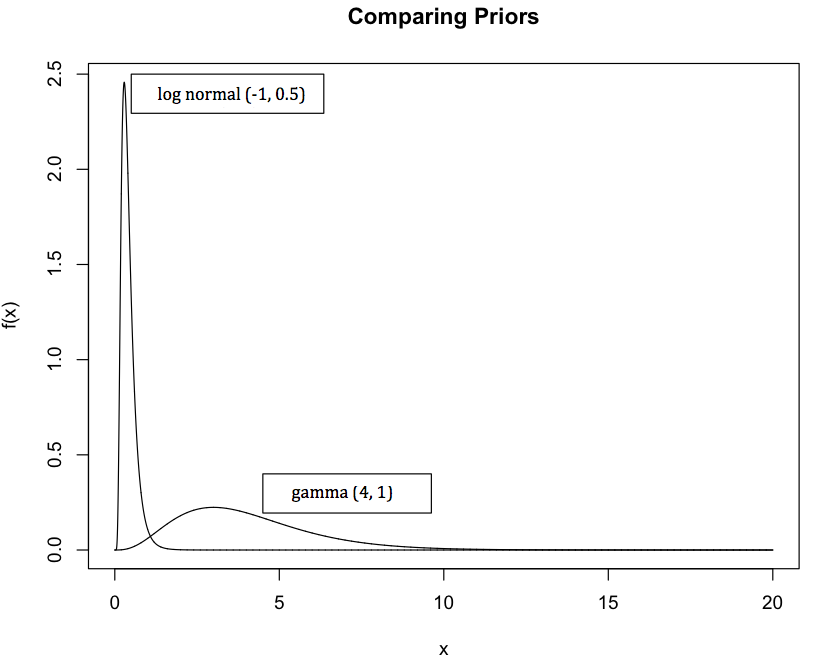}}
\caption{\small Comparison of lognormal and gamma priors for the standard deviation of the regression slope.} 
\label{COP}
\end{figure}

Consider for instance the variation between the choices of lognormal and gamma
distributions as priors of $S_2$, standard deviation of the regression slope.
Figure \ref{COP} compares the lognormal prior with mean $-1$ and standard
deviation $0.5$ and the gamma distribution ${G}(4, 1)$.

In this case, most of the mass of the lognormal prior is concentrated on values
close to zero whereas  the gamma prior is more diffuse.  The $10, 50, 90$
percentiles of $LN(-1, 0.5)$ and $G(4, 1)$ are $0.19, 0.37, 0.7$ and $1.74,
3.67, 6.68$, respectively. Thus,  choosing $LN(-1, 0.5)$ as the prior of $S_2$
is equivalent to believe that values of $ \beta_2$ in the $m$ regressions are
much closer together than the situation where we assume $S_2\sim G(4, 1)$. To
assess the difference between both prior choices on $S_2$ and their impact
on the degree of similarity of the regression coefficients, we resort to a
simulated  example, similar to \citet{Bmm}, except that $m=4$ and $n_j=36$.
The explanatory variables are simulated standard normal variates.
We also take $\tau_j \sim IG(3, 1)$ and $ \bar{\beta} \sim N(0,
1000I)$. The prior for $\Sigma$ is such that $\pi (R) \propto 1$ and we run
Seaman et al.'s (2012) analyses under $S_2\sim LN( -1, 0.5)$ and $S_2 \sim G(4,
1)$. 

\subsection{Comparison of posterior outputs}
\label{sec:six1}

\begin{figure}[!h]
\centerline{
\includegraphics[scale=1,width=1\textwidth]{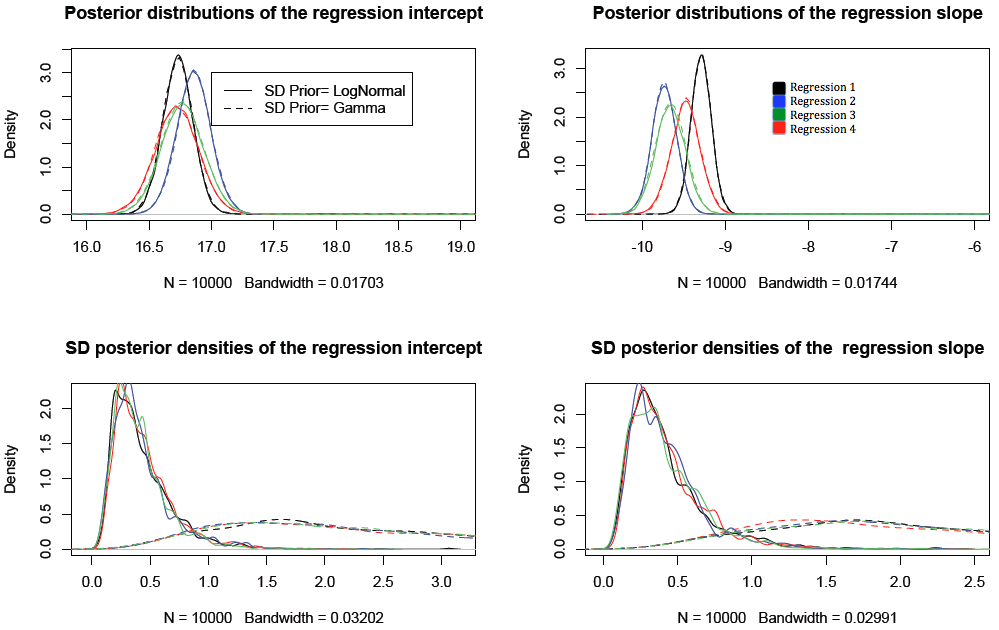}}
\caption{\small Estimated posterior densities of the regression intercept (top left), slope (top right), standard deviation of the intercept (down left) and standard deviation of the slope (down right), respectively for $4$ different normal regressions. All estimates based on $10^5$ iterations that were simulated from a Gibbs sampler.}
\label{cs}
\end{figure}

As seen in Tables \ref{jx} and \ref{jsd}, respectively. The differences
between the regression estimates are quite limited from one prior to the next,
while the estimates of the standard deviations vary much more.  In the
lognormal case, the posterior of   $S_i$ is concentrated on smaller values
relative to the gamma prior.  Figure \ref{cs} displays the posterior
distributions of  those parameters.
the impact of the prior choice
is quite clear on the standard deviations. Therefore, since the posteriors
of both intercepts and slopes for all four regressions are centered in $(16.5, 17)$ 
and $(-10, -9)$, respectively, we can conclude at the stability of Bayesian inferences on
$\beta_j$ when selecting two different prior distributions on $S_j$. That the posteriors
on the $S_i$'s differ is in fine natural since those are hyperparameters that are poorly informed by
the data, thus reflecting more the modelling choices of the experimenter.

\begin{table}
  \caption{ Posterior estimations of regression coefficients when their standard deviations 
are distributed as $LN(-1, 0.5)$ and $G(4, 1)$.  }
  \label{jx}
\centerline{
\begin{tabular}{|c|c|c|c|c|c|c|c|c|}
\hline
\multicolumn{9}{|c|}{$S_i \sim LN(-1, 0.5)$}  \\
\multicolumn{3}{|c|}{Regression 1} & \multicolumn{2}{|c|}{Regression 2} & \multicolumn{2}{|c|}{Regression 3} & \multicolumn{2}{|c|}{Regression 4} \\   
\hline  
 Estimate  & mean & s.d & mean & s.d & mean & s.d & mean & s.d \\ 
 Intercept &16.74 & 0.17 & 16.72 & 0.17&16.79 & 1.09&16.82&0.69\\
\hline 
 Slope      & -9.27 & 0.42 & -9.47 & 0.25& -9.66&0.98&-9.63&0.45 \\
\hline
\multicolumn{9}{|c|}{$S_i \sim G(4, 1)$} \\
\multicolumn{3}{|c|}{Regression 1} & \multicolumn{2}{|c|}{Regression 2} & \multicolumn{2}{|c|}{Regression 3} & \multicolumn{2}{|c|}{Regression 4}  \\   
\hline  
 Estimate  & mean & s.d & mean & s.d & mean & s.d & mean & s.d \\ 
 Intercept &16.73 & 0.23&16.73&0.22&16.85&0.37&16.76&0.32 \\
\hline 
 Slope      & -9.30 & 0.30&-9.47&0.34&-9.73&0.23&-9.64&0.80 \\
\hline
\end{tabular}}
\end{table}

\begin{table}
  \caption{ Posterior estimations standard deviations of the regression coefficients when their priors 
  are distributed as $LN(-1, 0.5)$ versus $G(4, 1)$.  }
  \label{jsd}
\centerline{
\begin{tabular}{|c|c|c|c|c|c|c|c|c|}
\hline
\multicolumn{9}{|c|}{$S_i \sim LN(-1, 0.5)$}  \\
\hline
\multicolumn{3}{|c|}{Regression 1} & \multicolumn{2}{|c|}{Regression 2} & \multicolumn{2}{|c|}{Regression 3} & \multicolumn{2}{|c|}{Regression 4} \\   
\hline  
 Estimate  & mean & s.d & mean & s.d & mean & s.d & mean & s.d \\ \cline{1-9}
 $S_1$          &0.43&0.27&0.44&0.26&0.42&0.26&0.41&0.24 \\
\hline 
 $S_2$          &0.42&0.27&0.43&0.25&0.42& 0.25& 0.43&0.32 \\
\hline
\multicolumn{9}{|c|}{$S_i \sim G(4, 1)$} \\
\hline
\multicolumn{3}{|c|}{Regression 1} & \multicolumn{2}{|c|}{Regression 2} & \multicolumn{2}{|c|}{Regression 3} & \multicolumn{2}{|c|}{Regression 4}  \\   
\hline  
 Estimate  & mean & s.d & mean & s.d & mean & s.d & mean & s.d \\ \cline{1-9}
 $S_1$          & 2.31&1.28& 2.33&1.29&2.29&1.29&2.29&1.26\\
\hline 
 $S_2$         & 2.32&1.29&2.23&1.28&2.25&1.23& 2.30&1.26\\
\hline
\end{tabular}}
\end{table}

\section{Examples 3 and 4: Prior choices for a proportion and the multinomial 
coefficients}
\label{sec:sept}

This section considers more briefly the third and fourth examples of
\citet{Hd}. The third example relates to a treatment effect analyzed by
\citet{Cw} and the fourth one covers a standard multinomial setting.

\subsection{Proportion of treatment effect captured}

In \citet{Cw} two models are compared for
surrogate endpoints, using a link function $g$ that either includes the
surrogate marker or not. The quantity of interest is a
proportion of treatment effect captured: it is defined as $\mathrm{PTE}
\equiv 1-{\beta_1}/{\beta_{R,1}}$, where $\beta_1, \beta_{R,1}$ are the
coefficients of an indicator variable for treatment in the first and second
regression models under comparison, respectively. \citet{Hd} restricted this proportion to the interval $(0,
1)$ and under this assumption they proposed to use a generalised beta distribution on
$\beta_1, \beta_{R,1}$ so that $\mathrm{PTE}$ stayed within $(0, 1)$. 

We find this example most intringuing in that,
even if PTE could be turned into a meaningful quantity (given that it depends
on parameters from different models), the criticism that it may take values outside
$(0,1)$ is rather dead-born since it suffices to impose a joint prior that ensures
the ratio stays within $(0,1)$. This actually is the solution eventually
proposed by the authors. If we have prior beliefs about the parameter space
(which depends on ${\beta_1}/{\beta_{R,1}}$ in this example) the prior
specified on the quantity of interest should integrate these
beliefs. In the current setting, there is seemingly no prior information about
$(\beta_1, \beta_{R,1})$ and hence imposing a prior restriction to
$(0, 1)$ is not a logical specification. For instance, using normal priors on
$\beta_1$ and $ \beta_{R,1}$ lead to a Cauchy prior on ${\beta_1}/{\beta_{R,1}}$,
which support is not limited to $(0, 1)$. We will not discuss this rather
artificial example any further.

\subsection{Multinomial model and evenness index}

\begin{figure}[!h]
\centerline{
\includegraphics[width=.7\textwidth]{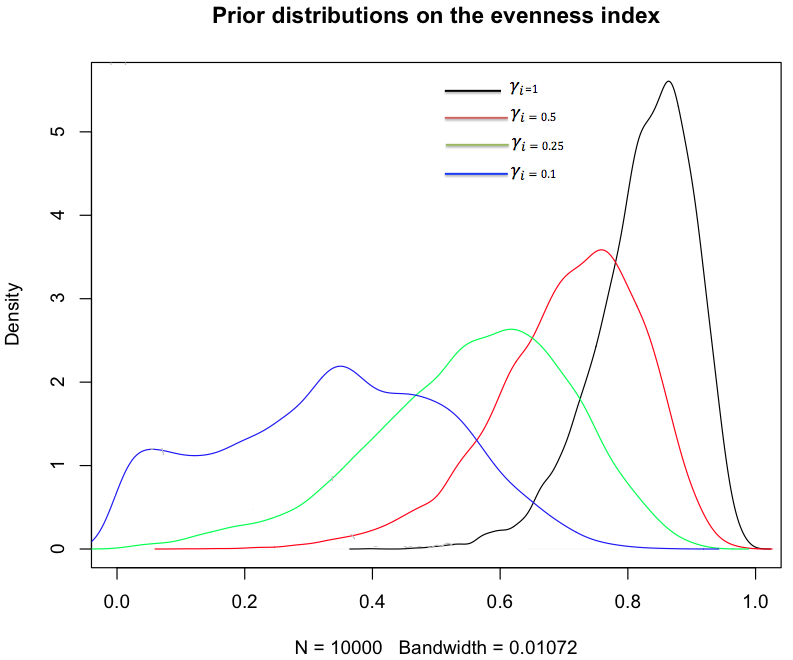}}
\caption{\small Priors induced on the evenness index: Four Dirichlet 
prior are assigned to $ \theta$ with hyperparameters
all equal to $0.1, 0.25, 0.5, 1$, based on $10^4$ simulations.}
\label{ph}
\end{figure}

\begin{figure}[!h]
\centerline{
\includegraphics[width=.7\textwidth]{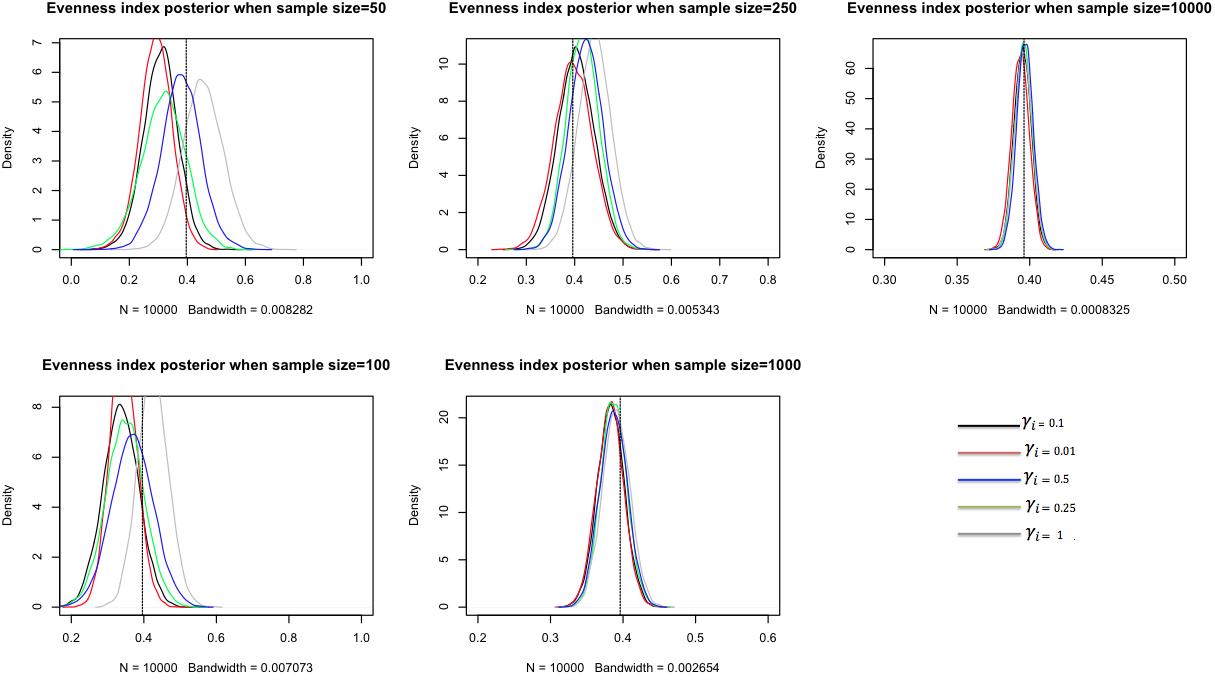}}
\caption{\small Estimated posterior densities of $H(\theta)$ considering sample sizes of $50, 100, 250, 1000, 10,000$. They correspond to the priors on $ \theta$ shown in Figure \ref{ph} and are based on $10^4$ posterior simulations. The vertical line indicates the mode of all posteriors when sample size is large enough. }
\label{pshl}
\end{figure}

The final example in \citet{Hd} deals with a measure 
called {\em evenness index} $H(\theta)={-\sum \theta_i \log(\theta_i)}
\big/{\log(K)}$ that is a function  of a vector
$\theta$ of proportions $\theta_i$, $i=1,\ldots,K$.
The authors assume a Dirichlet prior on $\theta$
with hyperparameters first equal to $1$ then to $0.25$. For the transform $H(\theta)$, 
Figure \ref{ph} shows that the first prior concentrates on $(0.5, 1)$ whereas the second does not.
Since there is nothing special about the uniform prior, re-running the
evaluation with the Jeffreys prior reduces this feature, which anyway is a
characteristic of the prior distribution, not of a posterior distribution
that would account for the data. The authors actually propose to use the
$\mathrm{Dir}(1/4,1/4,\ldots ,1/4)$ prior, presumably on the basis that the
induced prior on the evenness is then centered close to 0.5. If we consider the
more generic $\mathrm{Dir}(\gamma_1, \ldots, \gamma_K)$ prior, we can investigate the
impact of the $\gamma_i$'s when they move from $ 0.1$ to $1$.  In 
Figure \ref{ph}, the induced priors on $H(\theta)$ indeed show a decreasing
concentration of the posterior on $(0.5,1)$ as
$\gamma_i$ decreases towards zero.
To further the comparison, we generated datasets of size $N=50, 100, 250,
1000, 10,000$.  Figure \ref{pshl} shows the posteriors associated with each of
the four Dirichlet priors for these samples, including modes that 
are all close to $0.4$ when $N=10^4$. Even for moderate sample sizes like
$50$, the induced posteriors are almost similar. 
When the sample size is $50$, Table \ref{jxt} shows there is some degree of variation between the 
posterior means, even though, as expected, this difference vanishes when the sample size increases.

\begin{figure}[!h]
\centerline{
\includegraphics[width=.8\textwidth]{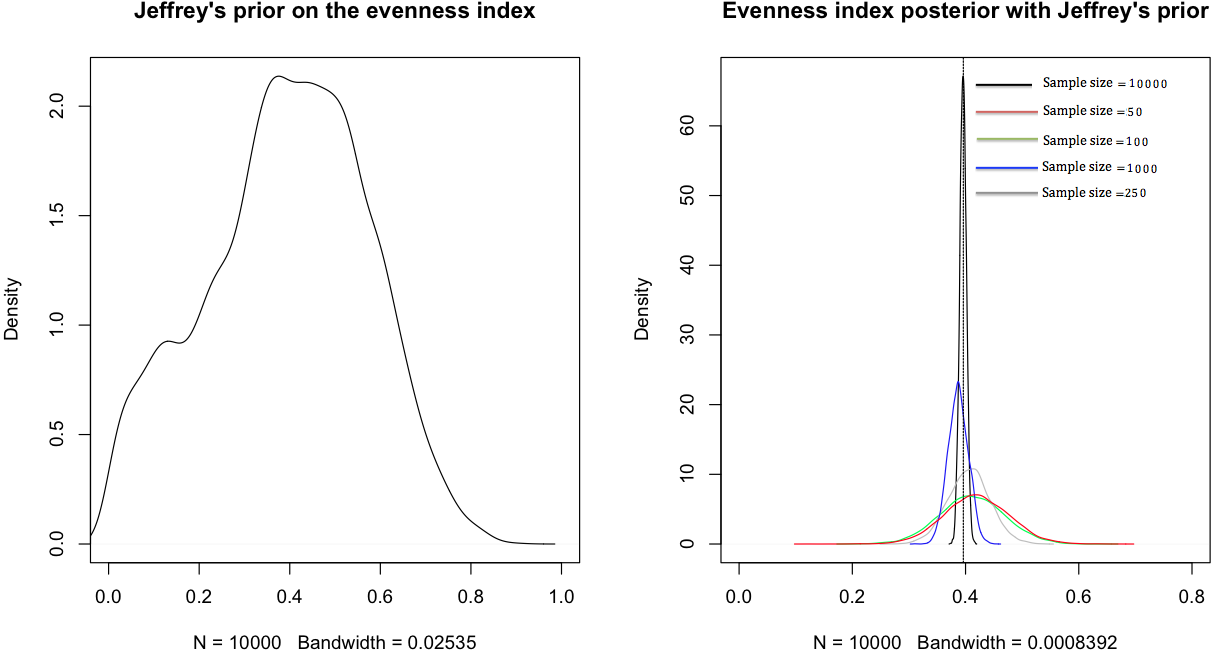}}
\caption{\small Jeffreys' prior and estimated posterior densities of $H(\theta)$ considering sample sizes $50, 100, 250, 1000, 10,000$. The posterior distributions are based on $10^4$ posterior draws. The vertical line indicates the mode of the posterior density when the sample size is $10^4$. }
\label{pjh}
\end{figure}

Note that, while Dirichlet distributions are conjugate priors, hence potentially lacking
in robustness, Jeffreys's prior is a special case corresponding to $\gamma_i=1/K$ (here $K$ is equal to $8$).
Figure \ref{pjh} reproduces the transform of Jeffreys' prior for the evenness
index (left) and the induced posterior densities for the same values of $N$.
Since it is a special case of the above, the same features appear. A potential alternative we did not
explore is to set a non-informative prior on the hyperparameters of the Dirichlet distribution.

\begin{table}
  \caption{ Posterior means of $H(\theta)$ for the priors shown in Figure \ref{ph} and Jeffreys' prior on $ \theta$ for sample sizes $50, 100, 250, 1000, 10,000$.  }
\label{jxt}
\centerline{
\begin{tabular}{|c|c|c|c|c|c|}
\hline
Sample size & 50 & 100 & 250 & 1000 & 10,000 \\ \cline{1-6}
\hline
\multicolumn{6}{|c|}{Dirichlet prior when $\gamma_i=0.1$}  \\
\hline
Posterior mean & 0.308 & 0.336 & 0.403 & 0.383 & 0.395 \\ \cline{1-6}  
\hline
\multicolumn{6}{|c|}{Dirichlet prior when $\gamma_i=0.25$}  \\
\hline
Posterior mean & 0.317 & 0.438 & 0.417 & 0.387 & 0.396 \\ \cline{1-6}  
\hline
\multicolumn{6}{|c|}{Dirichlet prior when $\gamma_i=0.5$}  \\
\hline
Posterior mean & 0.378 & 0.368 & 0.423 & 0.387 & 0.397 \\ \cline{1-6}  
\hline
\multicolumn{6}{|c|}{Dirichlet prior when $\gamma_i=1$}  \\
\hline
Posterior mean & 0.454 & 0.425 & 0.441 & 0.390 & 0.396 \\ \cline{1-6}  
\hline
\multicolumn{6}{|c|}{Jeffreys' prior: $\gamma_i=0.125$}  \\
\hline
Posterior mean & 0.413 & 0.411 & 0.406 & 0.390 & 0.396 \\ \cline{1-6}  
\hline
Posterior s.d  & 0.058 & 0.057 & 0.037 & 0.018 & 0.006 \\ \cline{1-6}   
\hline
\end{tabular}}
\end{table}

\vspace{0.3cm}
\section{Conclusion}\label{sec:quatr}

In this note, we have reassessed the examples supporting the critical review of \citet{Hd},
mostly showing that off-the-shelf noninformative priors are not suffering from the shortcomings
pointed out by those authors. Indeed, according to the outcomes produced therein, those
noninformative priors result in stable posterior inferences
and reasonable Bayesian estimations for the parameters at hand. We
thus consider the level of criticism found in the original paper rather unfounded,
as it either relies on a highly specific choice of a proper prior distribution or on
bypassing basic prior information later used for criticism. The paper of \citet{Hd} concludes with
recommendations for prior checks. These recommendations are mostly sensible if
mainly expressing the fact that some prior information is almost always
available on some quantities of interest. Our sole point
of contention is the repeated and recommended reference to MLE, if only because
it implies assessing or building the prior from the data. The most specific (if
related to the above)
recommendation is to use conditional mean priors as exposed by \citet{Crw}. For
instance, in the first (logistic) example, this meant putting a prior on the
cdfs at age $40$ and age $60$. The authors picked a uniform in both cases,
which sounds inconsistent with the presupposed shape of the probability
function.

In conclusion, we find there is nothing pathologically wrong with either the
paper of \cite{Hd} or the use of ``noninformative" priors! Looking at induced
priors on more intuitive transforms of the original parameters is a commendable
suggestion, provided some intuition or prior information is already available
on those. Using a collection of priors including reference or invariant priors
helps as well towards building a feeling about the appropriate choice or range of
priors and looking at the dataset induced by simulating from the corresponding
predictive cannot hurt.


\begin{thebibliography}{21}
\providecommand{\natexlab}[1]{#1}
\providecommand{\url}[1]{{#1}}
\providecommand{\urlprefix}{URL }
\expandafter\ifx\csname urlstyle\endcsname\relax
  \providecommand{\doi}[1]{DOI~\discretionary{}{}{}#1}\else
  \providecommand{\doi}{DOI~\discretionary{}{}{}\begingroup
  \urlstyle{rm}\Url}\fi
\providecommand{\eprint}[2][]{\url{#2}}

\bibitem[{Barnard et~al(2000)Barnard, McCulloch, and Meng}]{Bmm}
Barnard J, McCulloch R, Meng XL (2000) : {M}odeling {C}ovariance {M}atrices in
  {T}erms of {S}tandard {D}eviations and {C}orrelations with {A}pplication to
  {S}hrinkage. Statistica Sinica 10(4):1281--1312

\bibitem[{Berger(1980)}]{Br}
Berger JO (1980) : {S}tatistical {D}ecision {T}heory: {F}oundations,
  {C}oncepts, and {M}ethods. Springer-Verl., New York

\bibitem[{Berger(1984)}]{Rbv}
Berger JO (1984) : {T}he {R}obust {B}ayesian {V}iewpoint (with discussion).
  Robustness of Bayesian Analysis pp 63--144

\bibitem[{Bernardo and Smith(2009)}]{Bt}
Bernardo JM, Smith AF (2009) : {B}ayesian {T}heory. (Vol. 405), John Wiley \&
  Sons, New York

\bibitem[{Box and Tiao(2011)}]{Bto}
Box GE, Tiao GC (2011) : {B}ayesian {I}nference in {S}tatistical {A}nalysis.
  (Vol. 40), John Wiley \& Sons, New York

\bibitem[{Christensen et~al(2011)Christensen, Johnson, Branscum, and
  Hanson}]{Crw}
Christensen R, Johnson WO, Branscum AJ, Hanson TE (2011) : {B}ayesian {I}deas
  and {D}ata {A}nalysis: {A}n {I}ntroduction for {S}cientists and
  {S}tatisticians. CRC Press, Chapman \& Hall

\bibitem[{Cowles(2002)}]{Cw}
Cowles MK (2002) : {B}ayesian {E}stimation of the {P}roportion of {T}reatment
  {E}ffect {C}aptured by {S}urrogate {M}arker. Statistics in Medicine
  21(6):811--834

\bibitem[{Firth(1993)}]{Bme}
Firth D (1993) : {B}ias {R}eduction of {M}aximum {L}ikelihood {E}stimates.
  Biometrika 80(1):27--38

\bibitem[{Gelman et~al(2013)Gelman, Carlin, Stern, Dunson, Vehtari, and
  Rubin}]{Bad}
Gelman A, Carlin JB, Stern HS, Dunson DB, Vehtari A, Rubin DB (2013) :
  {B}ayesian {D}ata {A}nalysis. CRC press, Chapman \& Hall

\bibitem[{Jaynes(2003)}]{Jaynes}
Jaynes E (2003) Probability Theory. Cambridge University Press, Cambridge

\bibitem[{Jeffreys(1939)}]{Jeff}
Jeffreys H (1939) Theory of Probability, 1st edn. The Clarendon Press, Oxford

\bibitem[{Laplace(1820)}]{Lpl}
Laplace PS (1820) : {T}h{\'e}orie {A}nalytique des {P}robabilit{\'e}s.
  Courcier, Paris

\bibitem[{Liang et~al(2008)Liang, Paulo, Molina, Clyde, and Berger}]{Mgp}
Liang F, Paulo R, Molina G, Clyde MA, Berger JO (2008) : {M}ixtures of
  {G}-priors for {B}ayesian {V}ariable {S}election. The American Statistical
  Association 103(481):410--423

\bibitem[{Lopes and Tobias(2011)}]{Cp}
Lopes HF, Tobias JL (2011) : {C}onfronting {P}rior {C}onvictions: {O}n {I}ssues
  {P}rior {S}ensitivity and {L}ikelihood {R}obustness {B}ayesian {A}nalysis.
  Annu Rev of Economics 3(1):107--131

\bibitem[{Marin and Robert(2007)}]{Bco}
Marin JM, Robert CP (2007) : {B}ayesian {C}ore: {A} {P}ractical {A}pproach to
  {C}omputational {B}ayesian {S}tatistics. Springer

\bibitem[{Rissanen(2012)}]{Riss}
Rissanen J (2012) Optimal Estimation of Parameters. Cambridge University Press

\bibitem[{Robert(2007)}]{Bc}
Robert CP (2007) : {T}he {B}ayesian {C}hoice: From {D}ecision-{T}heoretic
  {F}oundations to {C}omputational {I}mplementation. Springer, Springer Science
  \& Business

\bibitem[{{Seaman III} et~al(2012){Seaman III}, Seaman~Jr, and Stamey}]{Hd}
{Seaman III} JW, Seaman~Jr JW, Stamey JD (2012) : {H}idden {D}angers of
  {S}pecifying {N}oninformative {P}riors. The American Statistician
  66(2):77--84

\bibitem[{Stojanovski and Nur(2011)}]{Psa}
Stojanovski E, Nur D (2011) : {P}rior {S}ensitivity {A}nalysis for a
  {H}ierarchical {M}odel. Proceeding of the Fourth Annual ASEARCH Conference pp
  64--67

\bibitem[{Welch and Peers(1963)}]{Wh}
Welch B, Peers H (1963) : {O}n {F}ormulae for {C}onfidence {P}oints based on
  {I}ntegrals of {W}eighted {L}ikelihoods. Journal of the R Statistical Society
  Ser B (Methodological) 25:318--329

\bibitem[{Zellner(1986)}]{Zr}
Zellner A (1986) : {O}n {A}ssessing {P}rior {D}istributions and {B}ayesian
  {R}egression {A}nalysis with {G}-{P}rior {D}istributions. In Bayesian
  Inference and Decis Tech: Essays in Honor of Bruno de Finetti 6:233--243

\end{thebibliography}

\end{document}